\documentclass[prd,twocolumn,floatfix,preprintnumbers]{revtex4}

\usepackage{graphicx}
\input{epsf}
\usepackage{epsf}
\usepackage{graphicx,epsfig}
\usepackage{bm}
\usepackage{latexsym,amssymb,amsmath,float}

\newcommand {\ga} {\ {\raise-.5ex\hbox{$\buildrel>\over\sim$}}\ }
\newcommand {\la} {\ {\raise-.5ex\hbox{$\buildrel<\over\sim$}}\ }

\def\be{\begin{equation}}
\def\ee{\end{equation}}
\def\ba{\begin{eqnarray}}
\def\ea{\end{eqnarray}}

\begin{document}

\title{Dark Radiation from Particle Decays during Big Bang Nucleosynthesis}
\author{Justin L. Menestrina and Robert J. Scherrer}
\affiliation{Department of Physics and Astronomy, Vanderbilt University,
Nashville, TN  ~~37235}

\begin{abstract}

Cosmic microwave background (CMB) observations suggest the possibility of an extra dark radiation component,
while the current
evidence from big bang nucleosynthesis (BBN) is more ambiguous.  Dark
radiation from a decaying particle can affect these two processes differently. 
Early decays add an additional radiation component to both the CMB and BBN,
while late decays can alter the radiation content seen in the CMB while
having a negligible effect on BBN.  Here we quantify this difference and
explore the intermediate regime by
examining particles decaying during BBN, i.e., particle lifetimes $\tau_X$
satisfying 0.1 $\rm{sec} < \tau_X < 1000 ~\rm{sec}$.  We calculate
the change in the effective number of neutrino species, $N_{eff}$,
as measured by the CMB, $\Delta N_{CMB}$, and the change in the effective
number of neutrino species as measured by BBN, $\Delta N_{BBN}$, as a function
of the decaying particle initial energy density and lifetime, where $\Delta
N_{BBN}$ is defined in terms of the number of additional two-component neutrinos
needed to produce the same change in the primordial $^4$He abundance
as our decaying particle.
As expected, for
short lifetimes ($\tau_X \la 0.1$ sec), the particles decay before the onset
of BBN, and $\Delta N_{CMB} = \Delta N_{BBN}$, while for long lifetimes
($\tau_X \ga 1000$ sec),
$\Delta N_{BBN}$ is dominated by the energy density of the nonrelativistic
particles before they decay, so that $\Delta N_{BBN}$ remains nonzero and
becomes independent of the particle lifetime.  By varying both
the particle energy density and lifetime, one can obtain any desired
combination of $\Delta N_{BBN}$ and $\Delta N_{CMB}$, subject to the constraint
that
$\Delta N_{CMB} \ge \Delta N_{BBN}$.
We present limits on the decaying particle parameters derived from observational
constraints on $\Delta N_{CMB}$ and $\Delta N_{BBN}$.

\end{abstract}

\maketitle

\section{Introduction}

Over the past decade, a ``standard-model" cosmology has emerged,
based, among other things, on precision measurements of the fluctuations
in the cosmic microwave background (CMB) \cite{WMAP7A,WMAP7B}, observations
of type Ia supernovae \cite{union08,hicken} and big bang nucleosynthesis (BBN)
(see Ref. \cite{steigman} for a recent review).
In the standard cosmological model, the density of the universe
is dominated at present by a cosmological constant ($\Lambda$)
and cold dark matter (CDM), corresponding respectively to roughly
70\% and 25\% of the total density, with the remaining $5\%$ in baryons.
The present-day radiation content of the universe is negligible in comparison,
but this radiation was
the dominant component at early times.

While the cosmological observations are generally consistent
with this standard model, there are a few unresolved issues.  One
of these involves the total radiation content of the universe.
The energy density of
the CMB is a simple function of the CMB temperature and is known to high
accuracy.  Similarly, given the observed number of light neutrinos ($N_\nu = 3$), one can calculate the neutrino
energy density.  (In fact, the ``effective" number of neutrinos, $N_{eff}$ is slightly
greater than 3 due to partial heating of the neutrinos in the early universe
by electron-positron annihilation.  Including these effects
yields $N_{eff} = 3.046$ \cite{Dolgov,mangano1}).
However, recent precision measurements of
the CMB fluctuations are best fit by larger
values of $N_{eff}$. (For a discussion of the effect of $N_{eff}$ on the CMB
fluctuations, see Refs. \cite{Bashinsky,Hou}). The seven-year data from the Wilkinson
Microwave Anisotropy Probe, combined with observations of baryon acoustic
oscillations (BAO) and measurements of the Hubble parameter, $H_0$,
give $N_{eff} = 4.34^{+0.86}_{-0.88}$ (68\% CL)
\cite{WMAP7B}.  Observations by the Atacama Cosmology Telescope combined
with BAO and $H_0$ give
$N_{eff} = 4.56 \pm 0.75$ (68\% CL) \cite{Dunkley}.  Recent results from the South Pole
Telescope combined with WMAP7, BAO, and $H_0$ give $N_{eff} = 3.86 \pm 0.42$ (68\% CL)
\cite{Keisler}.  An analysis using combined datasets in Ref. \cite{Arch}
yields $N_{eff} = 4.08^{+0.71}_{-0.68}$ (95\% CL).  Since this additional
radiation component cannot interact electromagnetically, it has
been dubbed ``dark radiation".

BBN is also quite sensitive to the total radiation content in the universe,
but here the evidence is more ambiguous.  Recent calculations of
the relic helium abundance by Izotov and Thuan \cite{izotov}
and by Aver, Olive, and Skillman \cite{aver} have reached opposite conclusions,
with the former arguing for an additional dark radiation component
and the latter concluding that the standard number of neutrinos suffices.
A recent analysis by Mangano and Serpico \cite{mangano2} gives $\Delta N_{eff}
\le 1$ (95\% CL).
(See also the discussion in Ref. \cite{steig}).  An overview
combining CMB and BBN constraints on dark radiation can be found in
Ref. \cite{Hamann}, while constraints on the physical properties of the
dark radiation are discussed in Refs. \cite{Arch,Smith}.

Given these hints of new physics, a number of models have been proposed
to account for additional dark radiation.  The simplest way to achieve
this is to add additional relativistic relic particles, as suggested by,
e.g., Refs. \cite{hamann,nakayama,smirnov,feng}.  In this case, the value
for $N_{eff}$ determined by BBN and the CMB should be the same.  Additional
relativistic energy density can also be provided by a neutrino chemical
potential, as in the models discussed in Refs. \cite{pastor,krauss}.

Alternately, if one wishes to produce an increase in $N_{eff}$ in the CMB,
but retain the standard-model value for $N_{eff}$ in BBN, then an obvious
possibility is the production of relativistic, non-electromagnetically-interacting particles
from the decay of a massive relic particle after BBN.  Such a scenario
for the dark radiation was considered
by Ichikawa et al. \cite{Ichikawa}, and further elaborated by Fischler and
Meyers \cite{Fischler} and Hasenkamp \cite{Hasenkamp}.
If we let $N_{BBN}$ denote the value of $N_{eff}$ as measured by BBN,
and $N_{CMB}$ be the value determined from the CMB and other low-redshift
measurements, then these decaying particle models predict
$N_{BBN} \ll N_{CMB}$, while additional stable relativistic degrees
of freedom give $N_{BBN} = N_{CMB}$.

In this paper, we fill in the gap between these two regimes by examining
decaying particle scenarios in which the particle decays during BBN.
Such models produce a relation between $N_{BBN}$ and $N_{CMB}$ that
varies from $N_{BBN} = N_{CMB}$ at short particle lifetimes (when the
particle decays before the onset of BBN) to $N_{BBN} \ll N_{CMB}$ when
the particle decays after the conclusion of BBN.  In the next section,
we give a detailed discussion of our calculation and present our results
for $N_{BBN}$ and $N_{CMB}$ as a function of the decaying particle abundance
and lifetime.  Our conclusions, including observational
limits, are discussed in Sec. III.

\section{Dark Radiation from a Decaying Particle}

We assume a standard flat Friedman-Robertson-Walker model with the expansion rate given by:
\begin{equation}
H = \frac{\dot{R}}{R} = \left(\frac{8}{3}\pi G\rho\right)^{1/2},
\end{equation}
where R is the scale factor and $\rho$ is the energy density.  To this
standard cosmological model we add a nonrelativistic particle $X$, which
is unstable and decays with lifetime $\tau_X$.  By assumption, $X$ decays only
into ``invisible" relativistic decay products, which do not interact electromagnetically
and can thus form the dark radiation (see Refs.
\cite{Ichikawa,Fischler,Hasenkamp} for examples of such models).  The
effects of such decays during BBN were previously considered by the authors
of Ref. \cite{InertDecay}, and our treatment closely follows theirs.  (For
another early discussion of BBN with such decays, see Ref.
\cite{Khlopov}.)
The main focus of
Ref. \cite{InertDecay} was the constraint from BBN that could be placed on these
decaying particles, while in this paper, we will be interested in using such models to
provide dark radiation, and to determine the relation between
$N_{BBN}$ and $N_{CMB}$ as a function of the model parameters.
\begin{figure*}
\begin{center}
	\epsfig{file=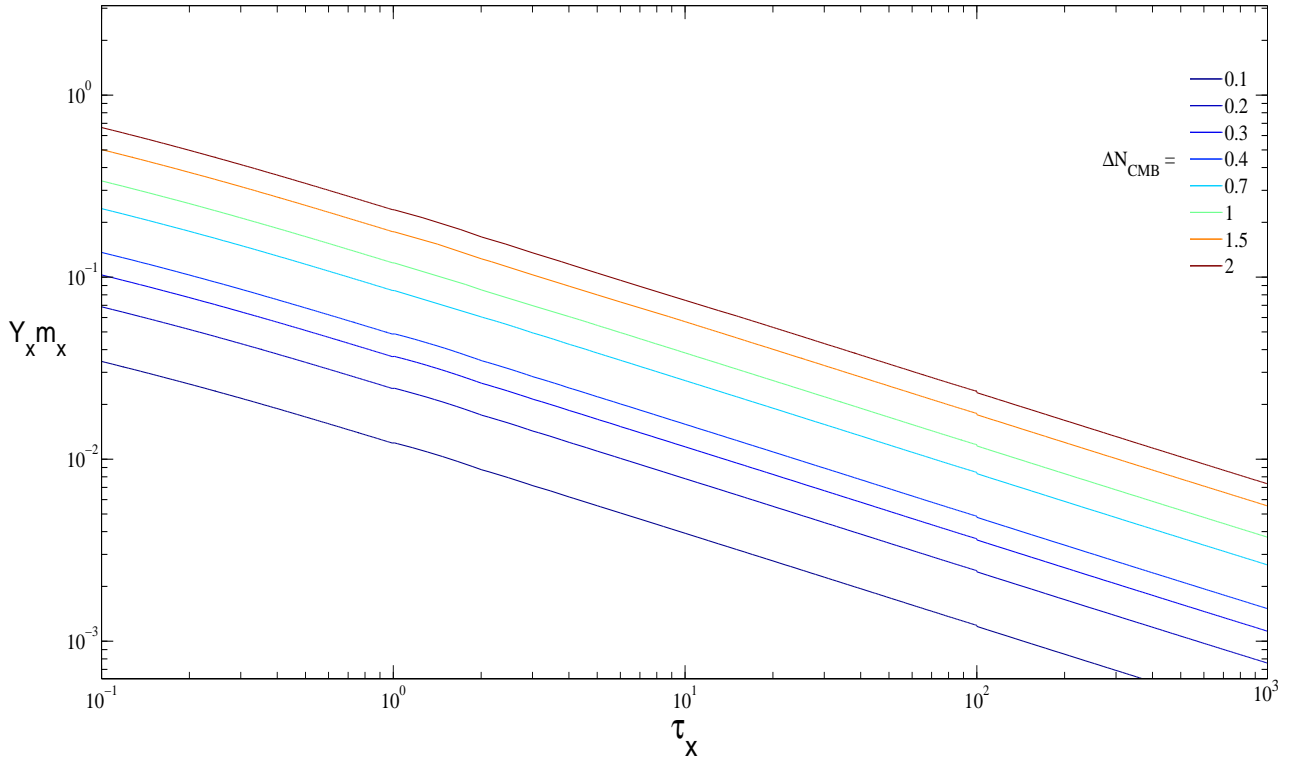,height=110mm, width = 200mm}
	\caption
	{Contour plot of $\Delta N_{CMB}$, the change
	in the effective number of neutrinos determined by
	CMB observations due to a decaying
	particle with lifetime $\tau_X$ and
	energy density prior to decay parametrized in terms of
	$Y_X m_X$, where $Y_X$ is the
	initial number density of the particle relative to entropy density, and
	$m_X$ is the particle mass.  Curves correspond to, from bottom
	to top, $\Delta N_{CMB} = 0.1, 0.2, 0.3, 0.4, 0.7, 1.0, 1.5, 2.0$.}
\end{center}
\end{figure*}

We follow Ref. \cite{KolbTurner} and parametrize the density of the decaying
particle in terms of its number density relative to the
entropy density, $s$, prior to decay ($t \ll \tau_X$):
\begin{equation}
Y_X = \frac{n_X}{s},
\end{equation}
where $s$ is given by
\begin{equation}
\label{entropy}
s = \frac{2 \pi^2}{45}g_{*s} T_\gamma^3.
\end{equation}
In Eq. (\ref{entropy}), $T_\gamma$ is the photon temperature, and
$g_{*s}$ is the effective number of ``entropy" degrees of freedom,
defined as
\begin{equation}
\label{gs}
g_{*s} = \sum_{bosons}g_B (T_i/T_\gamma)^3 + (7/8)\sum_{fermions} g_F
(T_i/T_\gamma)^3,
\end{equation}
where the sum is over all relativistic bosons and fermions.
In equation (\ref{gs}), $g_B$ and $g_F$ are the total number of
boson and fermion spin degrees of freedom, and $T_i$ is the temperature of
a given relativistic particle species.
The important useful property of this parametrization is that
$Y_X$ remains constant through the epoch of $e^+e^-$ annihilation, which
occurs during the middle of BBN.  During the BBN epoch, we have
$g_{*s} = 43/4$, and $n_X/n_\gamma = 19.36 ~Y_X$ prior to $e^+e^-$ annihilation,
while $n_X/n_\gamma = 7.04 ~Y_X$ after $e^+e^-$ annihilation.  For comparison,
the results of Ref. \cite{InertDecay} are parametrized in terms of $n_X/n_\gamma$
prior to $e^+e^-$ annihilation.

The equations governing the evolution of $\rho_X$ and the decay-produced
``invisible" radiation component, $\rho_{dec}$, are \cite{noheat}
\begin{eqnarray}
\label{drhoxdt}
\frac{d\rho_X}{dt} &=& -3H \rho_X - \rho_X/\tau_X, \\
\label{drhodecdt}
\frac{d\rho_{dec}}{dt} &=& - 4H \rho_{dec} + \rho_X/\tau_X.
\end{eqnarray}
Equation (\ref{drhoxdt}) gives
\begin{equation}
\rho_x =\rho_{x0}\left({R}/{R_0} \right)^{-3}e^{-t/\tau_X}, 
\end{equation}
while equation (\ref{drhodecdt}) must be integrated numerically
(although an analytic solution can be derived for $t \ll \tau_X$ \cite{noheat}).
It is clear from Eqs. (\ref{drhoxdt})-(\ref{drhodecdt}) that $\rho_{dec}$,
and thus, the increase in $N_{eff}$, depends only on $Y_X m_X$ and $\tau_X$.

Evolving these equations to calculate $\rho_X$ and $\rho_{dec}$, and converting
$\rho_{dec}$ into an effective number of neutrinos, we derive $\Delta N_{CMB}$, the change
in $N_{eff}$ as measured by the CMB, as a function of $Y_X m_X$ and $\tau_X$, where we confine
our attention to the case where the $X$ particle fully decays before
last scattering.  A contour plot of $\Delta N_{CMB}$ as a function of
$Y_X m_X$ and $\tau_X$ is given in Fig. 1.

In the limit where the decaying particles themselves never dominate
the expansion, one can calculate $\Delta N_{CMB}$ as a function
of $Y_X m_X$ and $\tau_X$, as in Ref. \cite{InertDecay}.  Rewriting
the results of Ref. \cite{InertDecay} in terms of our parameters, we find
\begin{equation}
\label{NCMB}
\Delta N_{CMB} = 8.3~(Y_X m_X/{\rm MeV})(\tau_X/{\rm sec})^{1/2}.
\end{equation}
A comparison of Eq. (\ref{NCMB}) with the results displayed in Fig. 1 shows
that Eq. (\ref{NCMB}) is accurate to within $\sim 10\%$ for the curves
displayed in Fig. 1.  Note that our analytic expression for $\Delta N_{CMB}$
differs significantly from that derived in Ref. \cite{Fischler}, as the latter
used the ``sudden decay approximation," in which all of the energy density
of the decaying particle is taken to be converted into relativistic decay
products at $t = \tau_X$.
\begin{figure*}
\begin{center}
	\epsfig{file=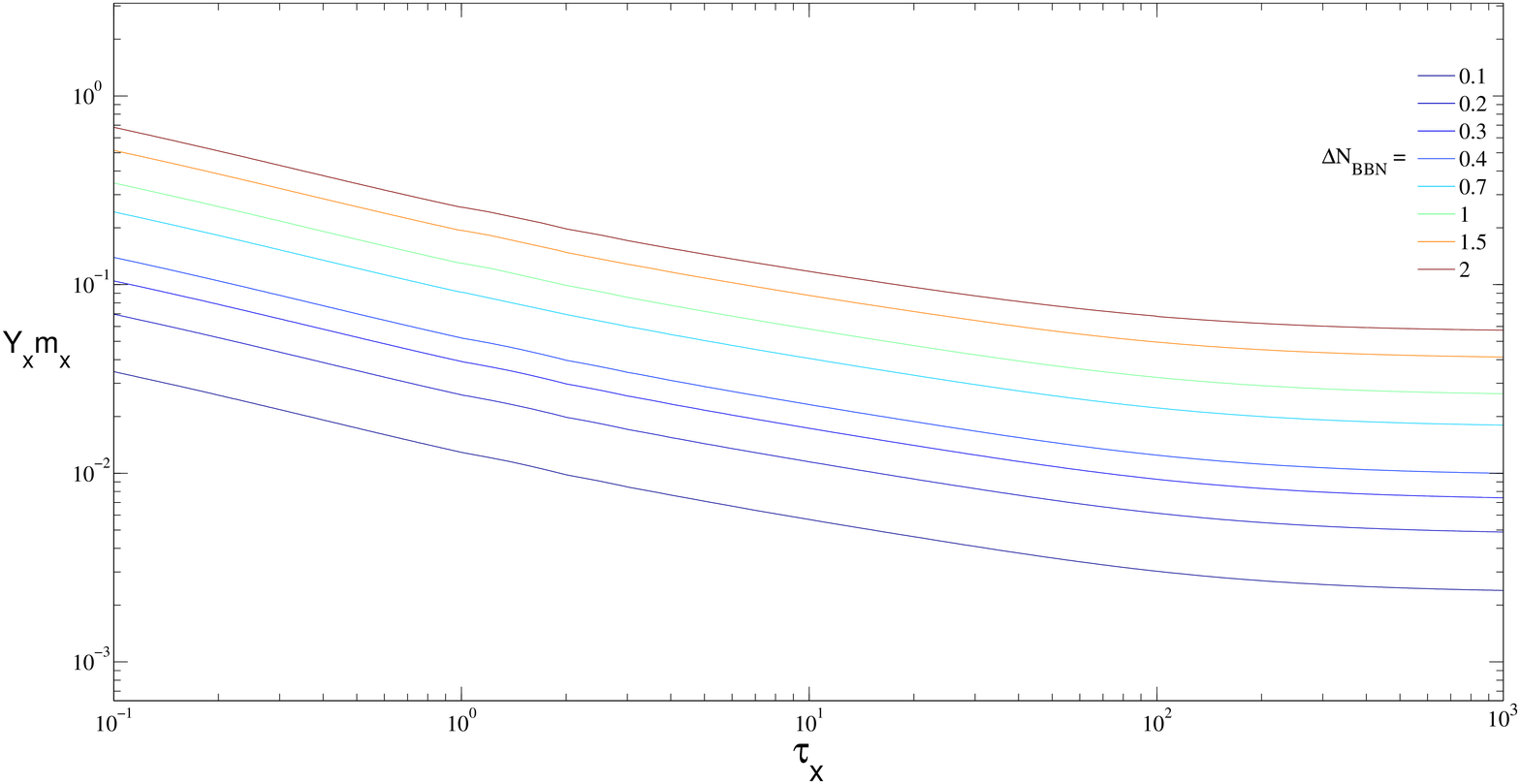,height=110mm,width = 200mm}
	\caption
	{Contour plot of $\Delta N_{BBN}$, the change
	in the effective number of neutrinos giving the same
	change in the primordial $^4$He abundance as a decaying particle with
	lifetime $\tau_X$ and energy density prior to decay
	parametrized in terms of
	$Y_X m_X$, where $Y_X$ is the
	initial number density of the particle relative to entropy density, and
	$m_X$ is the particle mass. Curves correspond to, from bottom to top,
	$\Delta N_{BBN} = 0.1, 0.2, 0.3, 0.4, 0.7, 1.0, 1.5, 2.0$.}
\end{center}
\end{figure*}

Adding additional energy density during BBN affects all of the element
abundances, but the effect is largest (relative to the accuracy with which
the primordial abundances can be estimated) for $^4$He.  Additional energy
density (either relativistic or nonrelativistic) increases the expansion rate,
resulting in a larger relic neutron abundance, and increased $^4$He production
\cite{steigman}.  Thus, the $^4$He abundance has long been used to constrain any
additional
relativistic energy density, such as that produced by additional
neutrinos \cite{SSG}.

We use the Kawano \cite{kawano} version
of the Wagoner \cite{wagoner1,wagoner2} Big Bang nucleosynthesis code
to calculate
the change in the primordial $^4$He abundance with the addition
of $\rho_X$ and $\rho_{dec}$ as given by Eqs. (\ref{drhoxdt}) and
(\ref{drhodecdt}), taking a baryon-photon ratio of
$\eta = 6.1 \times 10^{-10}$.  (Note that since we are
presenting the {\it change} in the $^4$He abundance
produced by the decaying particle,
rather than the absolute helium abundance itself,
our results are quite insensitive
to the assumed value of $\eta$; see, e.g., Fig. 6 of Ref. \cite{steigman}).
We
determine the change in the number of relativistic
neutrinos that gives exactly the same change
in the $^4$He abundance as a decaying particle
with a given abundance and lifetime.
Thus, for any pair of values for $Y_X m_X$ and
$\tau_X$, we have a corresponding change $\Delta N_{BBN}$ that
produces the same effect on BBN. (The change in the other element abundances
can be ignored here).
In Fig. 2, we give a contour plot of $\Delta N_{BBN}$ as a function
of $Y_X m_X$ and $\tau_X$.

\section{Conclusions}
In comparing Figs. 1 and 2, we see that for $\tau_X \la 0.1$ sec,
$\Delta N_{BBN} = \Delta N_{CMB}$.
In this short-lifetime limit, all of the decaying particle energy density
is converted into dark radiation before BBN
begins, so both BBN and the CMB ``see" the same $N_{eff}$.

In the opposite limit, $\tau_X \ga 1000$ sec, the contours
in Fig. 2 become horizontal lines.  In this long-lifetime limit,
all of the $X$ particles decay after BBN, and the increase
in the expansion rate that alters the $^4$He abundance is due entirely
to the energy density of the nonrelativistic particles before they decay.
Thus, in this
limit, $\Delta N_{BBN}$ becomes a function only of $Y_X m_X$ and is independent
of $\tau_X$.  Note that $\Delta N_{BBN}$ never goes to zero in the
long lifetime limit precisely because of this contribution to the expansion
rate from the nonrelativistc particles.
However, by increasing $\tau_X$, one can make
$\Delta N_{BBN}/\Delta N_{CMB}$ arbitrarily small.

The interesting transitional regime, then, is precisely the one we have
explored: 0.1 ${\rm sec} \la \tau_X \la 1000$ sec.
This would be the regime of interest if more precise measurements
of $N_{eff}$ from the CMB and BBN yielded nonzero values for both
$\Delta N_{CMB}$ and $\Delta N_{BBN}$ with $\Delta N_{BBN} \ne \Delta N_{CMB}$.
It this case, it is possible to simply read off, from Figs. 1-2,
values of $Y_X m_X$ and $\tau_X$ that give the desired values
for $\Delta N_{BBN}$ and $\Delta N_{CMB}$.  Note, however, that
one always has $\Delta N_{CMB} \ge \Delta N_{BBN}$ in this scenario,
so this model
can be falsified by observations contradicting this inequality.

\begin{figure*}
\begin{center}
	\epsfig{file=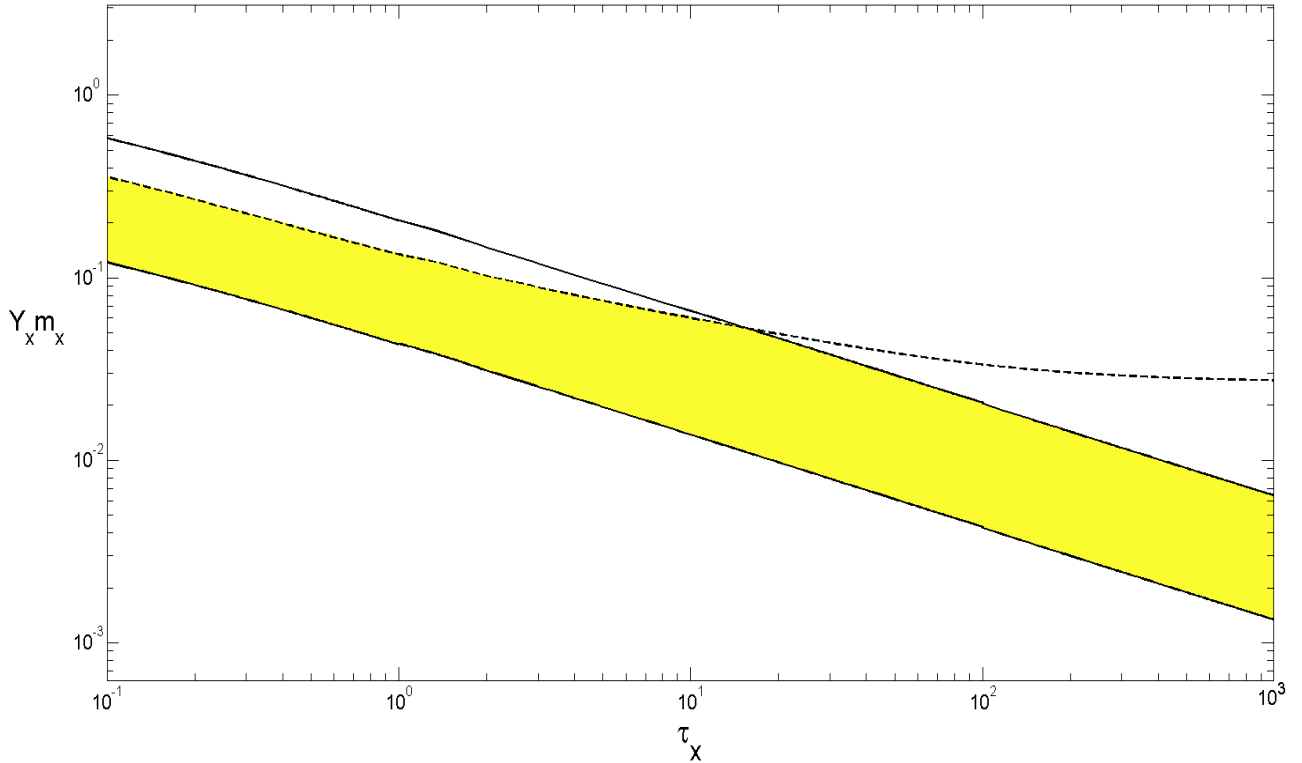,height=110mm,width = 200mm}
	\caption
	{Observational constraints from the CMB and BBN
	in the $\tau_X$, $Y_X m_X$ plane, where
	$\tau_X$ is the decaying particle lifetime,
	$Y_X$ is the
	initial number density of the particle relative to entropy density, and
	$m_X$ is the particle mass.  The region between the two solid
	curves is allowed by the CMB bounds on $N_{eff}$ from Ref. \cite{Arch},
	while the area below the dashed curve is the allowed region from BBN
	limits on $N_{eff}$
	given by Ref. \cite{mangano2}.  The shaded (yellow) region
	satisfies both constraints.}
\end{center}
\end{figure*}

As an example, we show, in Fig. 3, the limits on $Y_X m_X$ and
$\tau_X$ using the upper and lower bounds on $\Delta N_{CMB}$ from
Ref. \cite{Arch} and the upper bound on $\Delta N_{BBN}$ from
Ref. \cite{mangano2}.  As expected, the upper bound on $\Delta N_{BBN}$
cuts into the region favored by $\Delta N_{CMB}$ at short lifetimes, but current bounds are
not sufficiently restrictive
for this to be a major effect.  Tighter bounds
from future observational data will, of course, shrink this allowed region.

Our results can
be generalized to
more complicated scenarios,
such as a particle that decays into both
dark radiation and electromagnetically-interacting particles.
Even a small branching ratio into the latter can produce
a markedly different effect on BBN if the decay products are energetic
enough to photofission the primordial nuclei (see, e.g.,
Refs. \cite{kawasaki,jedamzik,kusakabe} and references therein).  These
effects, however, tend to be minimal for the lifetimes ($\tau_X \la 10^3$
sec) considered here, since the electromagnetically-interacting particles
thermalize rapidly at high temperatures.  Even in this case, however,
the electromagnetically-interacting decay products can heat the photons
relative to the neutrino background, potentially decreasing $N_{eff}$
instead of increasing it \cite{Fuller}.

\section{Acknowledgments}

R.J.S. was supported in part by the Department of Energy
(DE-FG05-85ER40226).

\end{document}